\documentclass[twoside,twocolumn,english,superscriptaddress,nofootinbib,preprintnumbers, aps]{revtex4}

\usepackage{euscript,amssymb}
\usepackage{amsthm}
\usepackage{graphicx}
\usepackage{amsfonts}
\usepackage{amsmath}
\usepackage{amssymb}
\usepackage{fancyhdr}
\usepackage{epsfig}
\usepackage{color}
\usepackage{esint}
\usepackage[unicode=true, pdfusetitle, bookmarks=true,bookmarksnumbered=false,bookmarksopen=false, breaklinks=false,pdfborder={0 0 1},backref=false,colorlinks=false]{hyperref}

\begin{document}

%%%%%%%%%%%%%%%%%%%%%%%%%%%%%%%%%%%%%%%%%%%%%%%%%%%%%%%%%%%%%%%%%%%%%%%%%%%%%%%%%%%%%
\title{Rainbow metric from quantum gravity}
%%%%%%%%%%%%%%%%%%%%%%%%%%%%%%%%%%%%%%%%%%%%%%%%%%%%%%%%%%%%%%%%%%%%%%%%%%%%%%%%%%%%%

\author{Mehdi Assanioussi}
\email{mehdi.assanioussi@fuw.edu.pl} \affiliation{Instytut Fizyki Teoretycznej, Uniwersytet Warszawski, ul. Pasteura 5, 02-093 Warszawa, Poland}

\author{Andrea Dapor}
\email{andrea.dapor@fuw.edu.pl} \affiliation{Instytut Fizyki Teoretycznej, Uniwersytet Warszawski, ul. Pasteura 5, 02-093 Warszawa, Poland}

\author{Jerzy Lewandowski}
\email{jerzy.lewandowski@fuw.edu.pl} \affiliation{Instytut Fizyki Teoretycznej, Uniwersytet Warszawski, ul. Pasteura 5, 02-093 Warszawa, Poland}

%%%%%%%%%%%%%%%%%%%%%%%%%%%%%%%%%%%%%%%%%%%%%%%%%%%%%%%%%%%%%%%%%%%%%%%%%%%%%%%%%%%%%
\begin{abstract}
In this letter, we describe a general mechanism for emergence of a rainbow metric from a quantum cosmological model. This idea is based on QFT on a quantum spacetime. Under general assumptions, we discover that the quantum spacetime on which the field propagates can be replaced by a classical spacetime, whose metric depends explicitly on the energy of the field: as shown by an analysis of dispersion relations, quanta of different energy propagate on different metrics, similar to photons in a refractive material (hence the name ``rainbow'' used in the literature). In deriving this result, we do not consider any specific theory of quantum gravity: the qualitative behavior of high-energy particles on quantum spacetime relies only on the assumption that the quantum spacetime is described by a wave-function $\Psi_o$ in a Hilbert space $\mathcal{H}_G$.
\end{abstract}
%%%%%%%%%%%%%%%%%%%%%%%%%%%%%%%%%%%%%%%%%%%%%%%%%%%%%%%%%%%%%%%%%%%%%%%%%%%%%%%%%%%%%

\date{\today}

%\pacs{???}

\maketitle

It has been argued \cite{emerg1,emerg2,emerg3,emerg4,emerg5} that classical gravity could be a collective phenomenon emerging from quantum degrees of freedom, not unlike fluid dynamics emerges from microscopic molecular interactions. It is often stated that such an effective spacetime should be described by a so-called ``rainbow metric" \cite{RAIN-intro1,RAIN-intro2}, i.e., a metric that depends somehow on the energy of the particles propagating on it: it is not difficult to conceive that probing such an effective spacetime with high enough energies leads eventually to corrections due to the underlying fundamental quantum structure.\footnote{
In condensed matter physics, it is well known that the propagation of photons in a refractive material can be described in terms of free photons on an energy-dependent metric.
}
A fundamental origin for rainbow metrics has been identified (in the principle of relative locality \cite{RAIN-theory}), their phenomenology has been studied \cite{RAINphenom1,RAINphenom2}, and tests (based on the Lorentz-violating nature of such energy-dependent metrics) have been proposed \cite{RAINtest1,RAINtest2}. What was missing until today -- as far as our knowledge goes -- is a general \emph{mechanism} which produces an emergent rainbow metric from a quantum spacetime. Indeed, while various proposals for quantum gravity\footnote
{
Prominent examples are loop quantum gravity \cite{LQG1,LQG2,LQG3}, string theory \cite{string} and causal dynamical triangulations \cite{CDT}
}
can all be argued to reproduce classical gravity in the low energy limit, it seems to us that a clean procedure to extract this limit is yet to be formulated.

In this letter, we put forward such a proposal. In Section 1, we provide a mechanism for emergence of cosmological spacetime from quantum cosmology in complete generality (we only require gravitational degree of freedom to be described in terms of a state $\Psi_o$ in a Hilbert space $\mathcal{H}_G$, and to be ``heavy'' compared to the matter degrees of freedom in the Born-Oppenheimer sense). The idea for this mechanism is based on QFT on quantum spacetime as first introduced in \cite{QT-FRW}. With no \emph{ad-hoc} input, we find that the effective metric describing the emergent spacetime is indeed of the rainbow type, as it depends on the wave-vector $k$ of the mode of the matter field.\footnote
{
It should be said that a second procedure exists to extract an emergent spacetime from such QFT on quantum spacetime [\cite{QT-pert1}], and it does not lead to a $k$-dependent metric. Whether the two approaches are physically equivalent is currently unknown, though the issue is being investigated by the authors.
}
In Section 2 we perform a low-energy expansion of this metric, and show that the first correction to the ``classical metric" $\bar{g}^o_{\mu \nu}$ is of order $\beta p^2/m^2$, where $p$ is the physical momentum of the mode, $m$ is the mass of the field, and $\beta$ is a simple function of $\Psi_o$. It is rather surprising that the only 
information needed to reconstruct the effective metric from the quantum spacetime is just parameter $\beta$. Finally, in Section 3, we study the modified dispersion relation of this emergent metric, and find that heavy particles -- as opposed to light ones -- behave in a different way than in classical gravity. In particular, the velocity of light remains an upper bound, but is now dependent on $\beta$.
\section{Effective metric for massive scalar field}
For definiteness, consider a scalar field $\phi$ of mass $m$ minimally coupled to gravity. Following the Hamiltonian treatment of linear perturbations in cosmology \cite{perturbacje1, perturbacje2}, we can separate the homogeneous and the inhomogeneous degrees of freedom, and the classical dynamics for mode $k$ of $\phi$ is generated (up to second order) by the Hamiltonian
\begin{align} \label{mat}
H_{k} = H_o - \dfrac{1}{2} H_o^{-1} \left[{\pi}_{k}^2 + (k^2 a^4 + m^2 a^6) {\phi}_{k}^2\right]
\end{align}
Here, $({\phi}_{k}, {\pi}_{k})$ are the phase space variables representing mode $k$, while $(a, \pi_a)$ are the conjugated variables representing the homogeneous degrees of freedom of gravity (that is, the scale factor and its momentum). $H_o$ is the part of the Hamiltonian that accounts for the evolution of these gravitational degrees of freedom at the same order.\footnote
{
All the remaining degrees of freedom of gravity do not affect $\phi_k$ at this order, and can thus be disregarded in light of the Born-Oppenheimer test field approximation (see later).
}

After formal quantization of matter and gravity, (\ref{mat}) defines the following Schroedinger-like equation
\begin{align} \label{q-mat}
-i \hbar \dfrac{d}{dt} \Psi = \left[\hat{H}_o - \dfrac{1}{2} \left(\hat{H}_o^{-1} \otimes \hat{{\pi}}_{{k}}^2 + \hat{\Omega}(k, m) \otimes \hat{{\phi}}_{{k}}^2\right)\right] \Psi
\end{align}
where
\begin{align} \label{geo-omega}
\hat{\Omega}(k, m) := k^2 \widehat{H_o^{-1} a^4} + m^2 \widehat{H_o^{-1} a^6}
\end{align}
and $\Psi \in \mathcal{H} = \mathcal{H}_G \otimes L_2(\mathbb{R}, d\phi_{{k}})$, with $\mathcal{H}_G$ being the Hilbert space of quantum gravity.\footnote
{
We do not specify anything about $\mathcal{H}_G$: \emph{any} theory of quantum gravity will do.
}
At this point, we take the \emph{test field} approximation: we assume that the scalar field does not back-react on the gravitational part. It is therefore allowed to retain only the 0th order in the Born-Oppenheimer expansion of $\Psi$: during the whole evolution $\Psi = \Psi_o \otimes \varphi$, where $\varphi \in L_2(\mathbb{R}, d\phi_{{k}})$ and $\Psi_o \in \mathcal{H}_G$ evolves via Schroedinger-like equation $-i d\Psi_o/dt = \hat{H}_o \Psi_o$. This being the case, we can trace away the gravitational part in (\ref{q-mat}) and obtain an equation for the matter part only:
\begin{align} \label{final}
i \hbar \dfrac{d}{dt} \varphi = \hat{H}_{{k}}^{\text{fun}} \varphi
\end{align}
where
\begin{align} \label{funH}
\hat{H}_{{k}}^{\text{fun}} := \dfrac{1}{2} \left[\langle\Psi_o | \hat{H}_o^{-1} | \Psi_o\rangle \hat{{\pi}}_{{k}}^2 + \langle\Psi_o | \hat{\Omega}(k, m) | \Psi_o\rangle \hat{{\phi}}_{{k}}^2\right]
\end{align}
The point first observed in \cite{QT-FRW} and further analysed in \cite{QT-BI,QT-pert1,QT-pert2,QT-bia}, is that equation (\ref{final}) resembles the Schroedinger equation for a quantum field $\phi$ on a suitably defined \emph{classical} spacetime. Let the spacetime be classically described by a metric $\bar{g}_{\mu \nu}$ of the Robertson-Walker type:
\begin{align} \label{eff-geo}
\bar{g}_{\mu \nu} dx^\mu dx^\nu = -\bar{N}^2 dt^2 + \bar{a}^2 (dx^2 + dy^2 + dz^2)
\end{align}
Construcing regular QFT on such a curved spacetime, one obtains for mode ${k}$ of $\phi$ the following effective Schroedinger equation:
\begin{align} \label{mass-eff-mat}
i \hbar \dfrac{d}{dt} \varphi = \hat{H}_{\vec{k}, m}^{\text{eff}} \varphi
\end{align}
where
\begin{align} \label{effH}
\hat{H}_{\vec{k}, m}^{\text{eff}} := \dfrac{1}{2} \left[\frac{\bar{N}}{\bar{a}^3} \hat{\pi}_{{k}}^2 + \dfrac{\bar{N}}{\bar{a}^3} (k^2 \bar{a}^4 + m^2 \bar{a}^6) \hat{\phi}_{{k}}^2\right]
\end{align}
In other words, we can replace the fundamental theory described by (\ref{q-mat}) with regular QFT on curved spacetime (\ref{eff-geo}), provided that the terms in the two Hamiltonians (\ref{funH}) and (\ref{effH}) match. This last requirement gives rise to a system of 2 equations for 2 unknowns:
\begin{align} \label{system}
\dfrac{\bar{N}}{\bar{a}^3} = \langle \hat{H}_o^{-1} \rangle, \ \ \ \ \ \dfrac{\bar{N}}{\bar{a}^3} (k^2 \bar{a}^4 + m^2 \bar{a}^6) = \langle \hat{\Omega}(k, m) \rangle
\end{align}
The solution of the system is
\begin{align} \label{sol-C}
\bar{N} = \bar{a}^3 \langle\hat{H}_o^{-1}\rangle, \ \ \ \ \ \bar{a} = \bar{a}(k^2/m^2)
\end{align}
where $\bar{a}(k^2/m^2)$ is the solution to the algebraic equation
\begin{align} \label{algeq}
\bar{a}^6 + \dfrac{k^2}{m^2} \bar{a}^4 - \delta = 0, \ \ \ \ \ \text{with} \ \delta := \dfrac{\langle\hat{\Omega}(k, m)\rangle}{m^2 \langle \hat{H}_o^{-1}\rangle}
\end{align}
It is a non-trivial fact that this equation has a unique positive solution for every $k \geq 0$. It is given explicitly by
\begin{align} \label{goodgod}
\bar{a}^2(k^2/m^2) = \left\{
\begin{array}{lllll}
u_+ + u_- - \dfrac{k^2}{3 m^2} & & & & \text{if} \ \dfrac{4k^6}{27m^6} \leq \delta
\\
\\
\dfrac{2 k^2}{3m^2} \cos \theta - \dfrac{k^2}{3m^2} & & & & \text{if} \ \dfrac{4k^6}{27m^6} > \delta
\end{array}
\right.
\end{align}
where
\begin{align} \label{params1}
u_\pm := \sqrt[3]{\dfrac{\delta}{2} - \dfrac{k^6}{27 m^6} \pm \sqrt{\dfrac{\delta^2}{4} - \dfrac{k^6}{27m^6}\delta}}
\end{align}
and
\begin{align} \label{params2}
\theta := \dfrac{1}{3}\arccos\left(-1 + \dfrac{27m^6}{2k^6} \delta\right)
\end{align}
The two functions in (\ref{goodgod}) match continuously at $k = k_o$, where $k_o$ is the unique positive solution to equation $\delta = 4k^6/27m^6$. For $k < k_o$ we are in the first case, while for $k > k_o$ in the second.
\section{Low-energy limit}
Let us expand (\ref{goodgod}) for $k \ll m$. Up to order $k^4/m^4$, we have
\begin{align} \label{low-k-lim}
\bar{a}^2\left(\frac{k^2}{m^2}\right) \approx \bar{a}^2_o \left[1 + \dfrac{\beta}{3} \left(\frac{k/\bar{a}_o}{m}\right)^2\right]
\end{align}
with
\begin{align} \label{coeff}
\bar{a}^2_o = \sqrt[3]{\frac{\langle \widehat{H_o^{-1} {a^6}} \rangle}{\langle \hat{H}_o^{-1} \rangle}}, \ \ \ \ \ \beta := \frac{\langle \widehat{H_o^{-1} {a^4}} \rangle}{\langle \hat{H}_o^{-1} \rangle^{1/3} \langle \widehat{H_o^{-1} {a^6}} \rangle^{2/3}} - 1
\end{align}
From $\bar{a}$ and $\bar{a}_o$ we find $\bar{N}$ and $\bar{N}_o$ via the first equation in (\ref{sol-C}). We can then identify two effective FLRW metrics: the low-energy one
\begin{align} \label{metric0}
\bar{g}^o_{\mu \nu} dx^\mu dx^\nu = -\bar{N}_o^2 dt^2 + \bar{a}_o^2 (dx^2 + dy^2 + dz^2)
\end{align}
and the $k$-dependent one
\begin{align} \label{metrick}
\bar{g}_{\mu \nu} dx^\mu dx^\nu = -\bar{N}^2 dt^2 + \bar{a}^2 (dx^2 + dy^2 + dz^2)
\end{align}
We can interpret $\bar{g}^o$ as the metric measured by a classical observer, while scalar field (and especially its relativistic modes) propagate on $\bar{g}(k)$.\footnote
{
There is no ambiguity in the definition of ``classical observer'' or ``low energy metric''. Indeed, since Lorentz symmetry is violated, there exists a preferred family of observers with respect to which statements such as $k \ll m$ are meaningful. We identify unambiguously this family as the cosmological (i.e., comoving) observers of the metric $\bar{g}^o$. The reason to consider such observers as ``classical'' is the following: if such an observer only performs measurement of geodesics of macroscopic bodies (for which $k \ll m$, $k$ being the momentum she measures), she will only investigate the regime in which $\bar{g} \approx \bar{g}^o$, therefore concluding that the spacetime is described by $\bar{g}^o$.
\\
A mathematical description of the situation is the following: there exists a manifold whose homogeneous and isotropic metric is $\bar{g}^o$; semiclassical observers coincide with cosmological observers of this metric; if the momentum $k$ of a particle as measured by such observers is non-negligible compared to the mass $m$, then such particle obeys a dispersion relation with $\bar{g}(k)$.
}

Suppose that an observer with 4-velocity $u^\mu$ detects a particle with 4-momentum $k_\mu$. The energy and (norm of) momentum measured by the observer are
\begin{align} \label{observedKIN}
E = u^\mu k_\mu = \dfrac{k_o}{\bar{N}_o}, \ \ \ p^2 = (\bar{g}_o^{\mu \nu} + u^\mu u^\nu) k_\mu k_\nu = \dfrac{k^2}{\bar{a}_o^2}
\end{align}
where we used the fact that $\bar{g}^o_{\mu \nu} u^\mu u^\nu = -1$ to discover that $u^\mu = (1/\bar{N}_o, 0, 0, 0)$. On the other hand, the particle satisfies the mass-shell relation in \emph{its} metric (\ref{metrick}):
\begin{align} \label{MassShell}
-m^2 = \bar{g}^{\mu \nu} k_\mu k_\nu = -\dfrac{k_o^2}{\bar{N}^2} + \dfrac{k^2}{\bar{a}^2} = -f^2 E^2 + g^2 p^2
\end{align}
having introduced the so-called ``rainbow functions'' \cite{RAIN-intro2}
\begin{align} \label{rainbowFunctions}
f := \dfrac{\bar{N}_o}{\bar{N}}, \ \ \ \ \ g := \dfrac{\bar{a}_o}{\bar{a}}
\end{align}
From (\ref{low-k-lim}) it is immediate to compute $f$ and $g$, which explicitly depend on the physical momentum $p = k/\bar{a}_o$:
\begin{align} \label{FandG}
f^2 = \left(1 + \dfrac{\beta}{3} \frac{p^2}{m^2}\right)^{-3}, \ \ \ \ \ g^2 = \left(1 + \dfrac{\beta}{3} \frac{p^2}{m^2}\right)^{-1}
\end{align}
We thus obtain a modified dispersion relation from (\ref{MassShell}), 
\begin{align} \label{dispREL}
E^2 = \dfrac{1}{f^2} \left(g^2 p^2 + m^2\right) \approx m^2 + (1 + \beta) p^2 + O(p^4)
\end{align}
As expected, the standard dispersion relation $E = m$ is recovered in the limit $p \ll m$. The first correction in the case $p \approx m$ is precisely $\beta$, a quantity of exquisitely quantum gravitational origin. Note that -- contrary to the general belief -- no particular role is played by Planck energy, $E_\text{Pl} \approx 10^{28} \ eV$. In fact, for a highly quantum spacetime we have $\beta \approx 1$, and hence the particles probe the quantum structure of spacetime already at $p \approx m$. For a proton, this would correspond to mild energies of order $10^9 \ eV$. On the other hand, it is clear that $\beta \approx 0$ for semiclassical states, and hence quantum gravity corrections are irrelevant for low-energy particles. We should mention that a similar result was recovered in the semiclassical limit in \cite{smolin}.
\section{Analysis and Discussion}
Having the dispersion relation, it is possible to compute the velocity of the mode:
\begin{align} \label{velocity}
v = \dfrac{dE}{dp} = \dfrac{1 + \beta}{\sqrt{m^2 + (1 + \beta) p^2}} p
\end{align}
This expression only holds in the limit $p \ll m$, but it is enough to show the deformation already at low energies. The exact dispersion relation (obtained numerically from (\ref{goodgod})) is represented in figures 1 and 2, where the classical one ($\beta \approx 0$) is compared with the choice
\begin{align} \label{quantum-state}
\dfrac{\langle \widehat{H_o^{-1} a^6} \rangle}{\langle \hat{H}_o^{-1} \rangle} = 0.9 \langle \hat{a}^3 \rangle^2, \ \ \ \ \ \dfrac{\langle \widehat{H_o^{-1} a^4} \rangle}{\langle \hat{H}_o^{-1} \rangle} = 1.1 \langle \hat{a}^3 \rangle^{4/3}
\end{align}
corresponding to $\beta \approx 0.2$ (a highly non-classical situation). In figure 2 we also show the dispersion relation for massless particles: there is no dependence of velocity of light on $p$ (though there is still a dependence on $\beta$, and hence on time). To see why this happens, consider the limit $k \gg m$ of (\ref{goodgod}). It leads to
\begin{align} \label{high-energy-metric}
\bar{a}^2\left(\dfrac{m}{k}\right) = \bar{a}_\infty^2 \left[1 + O\left(\dfrac{m^2}{k^2}\right)\right], \ \ \ \bar{a}_\infty^2 := \sqrt{\dfrac{\langle \widehat{H_o^{-1} a^4} \rangle}{\langle \hat{H}_o^{-1} \rangle}}
\end{align}
The zeroth order coincides with the solution to system (\ref{system}) with $m = 0$, that is, the system we would have obtained if we considered a massless scalar field from the start. The massless scalar field case was first studied in \cite{QT-FRW}, and $\bar{a}_\infty$ of (\ref{high-energy-metric}) coincides with the result therein. Now, since $\bar{a}_\infty$ is independent of $k$, the metric seen by particles with $m = 0$ is $k$-independent (though different from the semiclassical metric, $\bar{a}_o$). It is therefore not surprising that no mode-dependence is found in the velocity of light particles. This can be made explicit by computing the dispersion relation for such field: plugging $f = \bar{N}_o/\bar{N}_\infty$ and $g = \bar{a}_o/\bar{a}_\infty$ in the general formula (\ref{dispREL}) with $m = 0$, we derive
\begin{align}
E = \dfrac{\bar{a}^2_\infty}{\bar{a}^2_o} \ p = \sqrt{\dfrac{\langle \widehat{H_o^{-1} a^4} \rangle}{\langle \hat{H}_o^{-1} \rangle}} \sqrt[3]{\frac{\langle \hat{H}_o^{-1} \rangle}{\langle \widehat{H_o^{-1} {a^6}} \rangle}} \ p = \sqrt{1 + \beta} \ p
\end{align}
Thus, as far as massless particles are concerned, the effect of the quantum background amounts to a constant shift in the velocity of light. While this effect cannot be detected by any local measurement, we stress that the derivation was done for a massless scalar field, and might well be different in the case of photons.

\begin{figure}[t]
\begin{centering}
\includegraphics[height=2.2in]{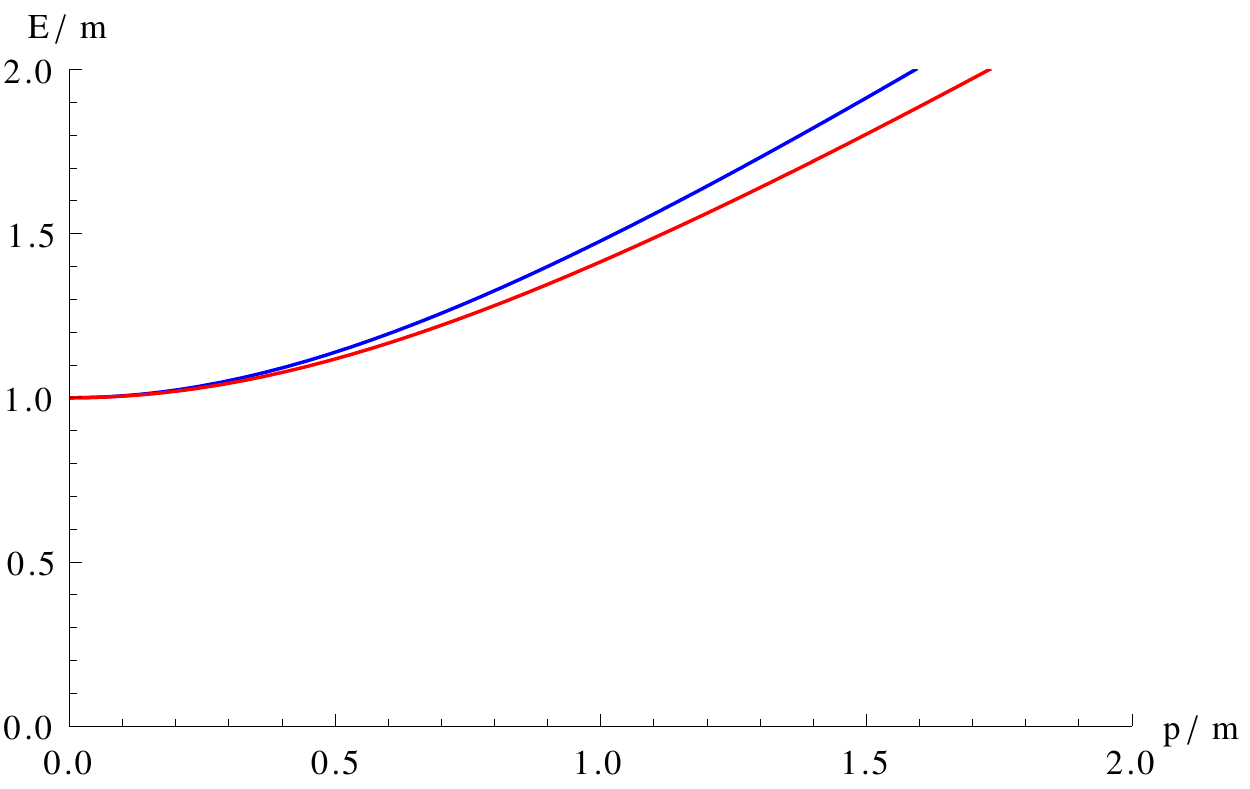}
\caption{{\footnotesize Dispersion relation $E = E(p)$ for a scalar field of mass $m$. Red = classical spacetime; Blue = quantum spacetime (eq. (\ref{quantum-state})).}}
\par\end{centering}
\label{fig:vels1}
\end{figure}

\begin{figure}[b]
\begin{centering}
\includegraphics[height=2.2in]{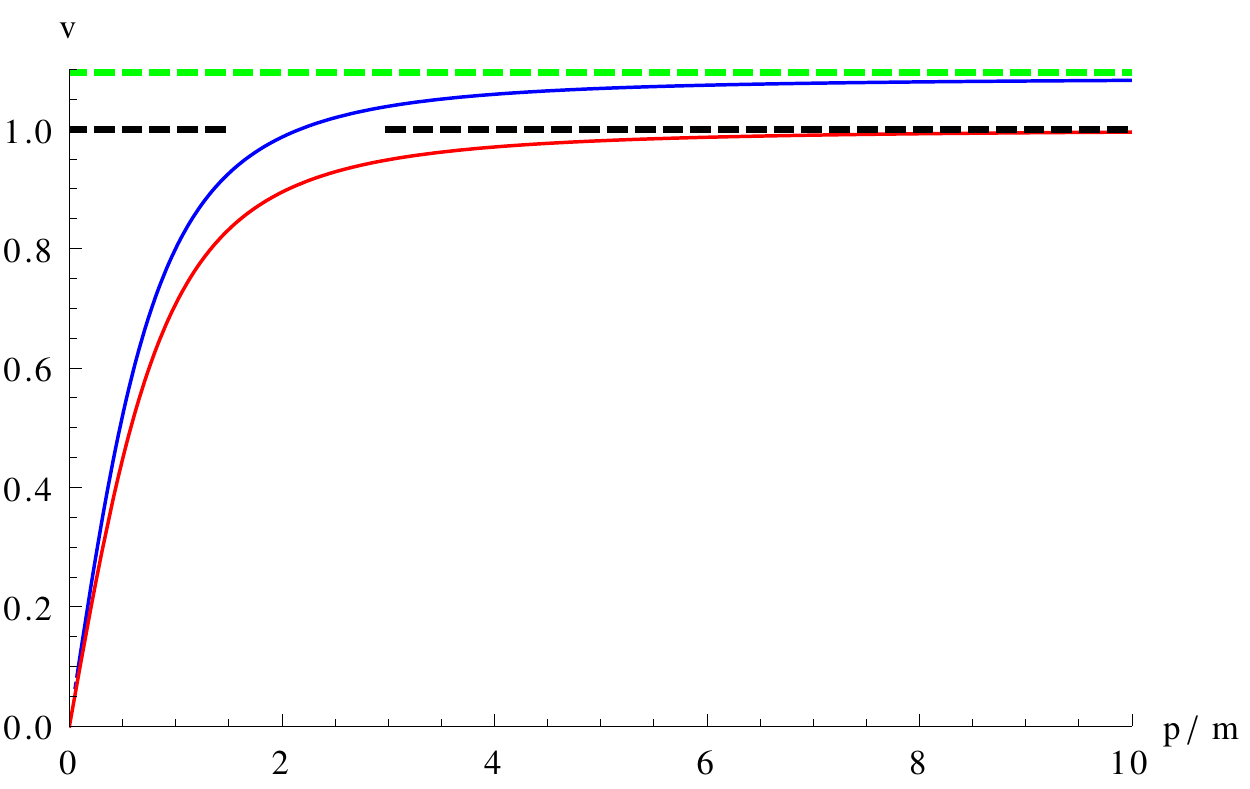}
\caption{{\footnotesize Velocity $v = v(p)$ of different modes of the massive field. Red = classical spacetime; Blue = quantum spacetime (eq. (\ref{quantum-state})). The dashed lines represent the speed of light in the semiclassical spacetime (black) and in the quantum spacetime (green) respectively.}}
\par\end{centering}
\label{fig:vels2}
\end{figure}

In conclusion, our analysis shows that the quantum nature of cosmological spacetime unavoidably affects the propagation of test particles, producing (apparent) Lorentz-violating effects. This has been shown by constructing an effective metric from general assumptions (in particular, we did not need to restrict to a specific quantum theory of cosmology). Intuitively, the generality of the result can be understood by observing that -- independently of the chosen quantum theory -- the quantum state $\Psi_o$ of the homogeneous gravitational field is not an eigenstate of the ``scale factor operator'' in general. Hence, it is to be expected that matter particles of different momenta $p$ will couple differently to the quantum geometry, probing different aspects of it. In the present work, we have used a massive scalar field $\phi$ -- one of the simplest forms of matter -- but there is no reason not to expect the same qualitative behaviour for other species.

As for the characterization of this effect, we have shown that the only parameter governing the corrections is $\beta$, a single function of $\Psi_o$. This is rather striking, considering that infinitely many quantum states can be found that give the same value $\beta$. On the other hand, we should not be too surprised, since the same happens for photons propagating in refractive media: the scattering of light through a crystal is perfectly well described in terms of the refractive index $n$, a single parameter in spite of the infinitely many possible microscopic configurations of the atoms.
Moreover, we should stress that the current results are based on a homogeneous quantum background: it is possible that inclusion of inhomogeneities might invalidate this result, but it is also conceivable that it will make it even more interesting (for instance, $\beta$ might depend on the position, as does the refractive index of a non-homogeneous medium).

But how strong is this effect? How big is $\beta$? While there is no \emph{fundamental} reason why $\beta$ should be small, it is an \emph{observational fact} that the universe today is classical. This means that the state $\Psi_o$ describing the current geometry of spacetime must be a coherent state with $\beta$ extremely small \cite{RAINtest1, RAINtest2}. Our philosophy is therefore to use $\beta$ as a test for the soundness of coherent states within a specific quantum cosmology or -- in case no good coherent state can be found -- as an indicator that said theory of quantum cosmology is incorrect.

While $\beta \approx 0$ today, several quantum gravity and quantum cosmology theories maintain that in the early stages of its life, the Universe should be described by a non-classical state. In this case $\beta \approx 1$, and the modified dispersion relations have to be taken into account when studying the behaviour of primordial matter. We can expect that traces of such effects are left on the CMB, or even speculate that they might have played a role in inflation and subsequent formation of structures.

%%%%%%%%%%%%%%%%%%%%%%%%%%%%%%%%%%%%%%%%%%%%%%%%
\section*{Acknowledgments}
%%%%%%%%%%%%%%%%%%%%%%%%%%%%%%%%%%%%%%%%%%%%%%%%

This work has been partially financed by the grants of Polish Narodowe Centrum Nauki nr 2011/02/A/ST2/00300 and nr 2013/09/N/ST2/04312.

%%%%%%%%%%%%%%%%%%%%%%%%%%%%%%%%%%%%%%%%%%%%%%%%%%%%%%%%%%%%%%%%%%%%%%%%%%%%%%%%%%%%%

\end{document}